# Observation of triplet superconductivity in CoSi$_2$/TiSi$_2$ heterostructures

Shao-Pin Chiu[1,2], C. C. Tsuei[3,4], Sheng-Shiuan Yeh[1,2,5], Fu-Chun Zhang[6], Stefan Kirchner[7,8,*], and Juhn-Jong Lin[1,2,9,*]

[1]Institute of Physics, National Chiao Tung University, Hsinchu 30010, Taiwan
[2]Center for Emergent Functional Matter Science, National Chiao Tung University, Hsinchu 30010, Taiwan
[3]IBM Thomas J. Watson Research Centers, Yorktown Heights, New York 10598, United States
[4]Institute of Physics, Academia Sinica, Nankang, Taipei 11529, Taiwan
[5]International College of Semiconductor Technology, National Chiao Tung University, Hsinchu 30010, Taiwan
[6]Kavli Institute for Theoretical Sciences and CAS Center for Excellence in Topological Quantum Computation, University of Chinese Academy of Sciences, Beijing 100190, China
[7]Zhejiang Institute of Modern Physics & Department of Physics, Zhejiang University, Hangzhou 310027, China
[8]Zhejiang Province Key Laboratory of Quantum Technology and Device, Zhejiang University, Hangzhou 310027, China
[9]Department of Electrophysics, National Chiao Tung University, Hsinchu 30010, Taiwan
[*]email: stefan.kirchner@correlated-matter.com; jjlin@mail.nctu.edu.tw

**Unconventional superconductivity and in particular triplet superconductivity have been front and center of topological materials and quantum technology research. Here we report our observation of triplet superconductivity in nonmagnetic CoSi$_2$/TiSi$_2$ heterostructures on silicon. CoSi$_2$ undergoes a sharp superconducting transition at a critical temperature $T_c \approx 1.5$ K, while TiSi$_2$ is a normal metal. We investigate conductance spectra of both two-terminal CoSi$_2$/TiSi$_2$ tunnel junctions and three-terminal T-shaped CoSi$_2$/TiSi$_2$ superconducting proximity structures. We report an unexpectedly large spin-orbit coupling in CoSi$_2$ heterostructures. Below $T_c$, we observe (1) a narrow zero-bias conductance peak on top of a broad hump, accompanied by two symmetric side dips in the tunnel junctions, (2) a narrow zero-bias conductance peak in T-shaped structures, and (3) hysteresis in the junction magnetoresistance. These three independent and complementary observations are indicative of chiral $p$-wave pairing in CoSi$_2$/TiSi$_2$**

**heterostructures. This chiral triplet superconductivity and the excellent fabrication compatibility of CoSi$_2$ and TiSi$_2$ with present-day silicon integrated-circuit technology facilitate full scalability for potential use in quantum-computing devices.**

Ever since the discovery that superfluid $^3$He realizes chiral *p*-wave pairing[1,2], condensed matter and materials scientists have been on the search for electronic systems that show triplet superconductivity[3-7]. This interest further intensified when it became clear that vortex cores of topological superconductors may offer a route towards the realization of non-Abelian statistics and fault-tolerant quantum computing[4,6,8,9]. Although several superconductors appear to be triplet superconductors, only a rather limited number is suspected to realize chiral *p*-wave pairing[3,5,10]. Superconductivity induced in the doped topological insulators[11,12] (e.g., Cu$_x$Bi$_2$Se$_3$ and Sn$_{1-x}$In$_x$Te), topological semimetals[13] (e.g., Cd$_3$As$_2$), and semiconductors with strong Rashba spin-orbit coupling[14,15] were recently discussed in this context.

Identifying chiral *p*-wave superconductors is challenging[5,10,16]. In superconductor/normal metal (S/N) junctions, the paring potential can be probed by superconducting tunneling spectroscopy which is essentially phase sensitive[17-22]. T-shaped proximity structures (see below, Fig. 3a), on the other hand, are anticipated to display a zero-bias conductance peak (ZBCP) for spin-triplet superconductors[23], while spin-singlet superconductors result in conductance dips. This kind of conductance dip was indeed observed in T-shaped aluminum-copper proximity structures despite the technical challenges to fabricate such T-shaped junctions[24].

Here we report successful fabrication of and systematic measurements on these two types of heterostructures using CoSi$_2$ as S and TiSi$_2$ as N components. The conductance spectroscopies based on these two types of devices provide independent and complementary evidences for the unexpected observation of a dominant triplet pairing amplitude in CoSi$_2$/TiSi$_2$ heterostructures. In the superconducting state of CoSi$_2$, i.e., for $T < T_c \simeq 1.5$ K, we observe a two-step ZBCP consisting of a broad hump and a sharp peak in the S/N tunnel junctions as well as a sharp ZBCP in the T-shaped S/N proximity structures. In our design, the sizes of all involved devices are such that they give rise to substantial Thouless energies which determine the full-width at half-maximum (FWHM) of the proximity-induced ZBCPs[23].

Paramagnetic CoSi$_2$ is a type II superconductor with a fairly small Ginzburg-Landau parameter $\kappa = \lambda / \xi \approx 0.82 > 1/\sqrt{2}$ and a correspondingly large $B_{c1} / B_{c2}$ (S3), where $\lambda$ is the penetration depth, $\xi$ is the coherence length, and $B_{c1}$ and $B_{c2}$ are the lower and upper critical magnetic fields, respectively. TiSi$_2$ is a non-magnetic metal. Three-dimensional TiSi$_2$ wires show no superconducting signal down to 50 mK (S4). A highly transparent, clean CoSi$_2$/TiSi$_2$ interface embedded in a Si(100) substrate (Figs. 1b and c, Methods, S1 and S2) plays a key role in our study. Due to their epitaxial growth characteristics, our CoSi$_2$ films (105 nm thick in all cases) possess a long lateral electron mean free path ($l_e \approx 300$ nm) and ultralow $1/f$ noise[25]. We note that the $1/f$ noise magnitudes of epitaxial CoSi$_2$/Si(100) films are about two to three orders of magnitude lower than those in polycrystalline (single-crystalline) aluminum films[25] which are commonly used as a basis for superconducting qubits and quantum computing. Moreover, high-resolution transmission electron microscopy (TEM) studies have indicated that the as-grown epitaxial CoSi$_2$/Si(100) heterostructure is extremely clean, without any detectable traces of residual (magnetic) Co or (nonmagnetic) CoSi and Co$_2$Si after thermal annealing[25]. The absence of magnetic impurities is confirmed by the SQUID magnetization measurements (S12) and further supported by our measured $T_c \simeq 1.5$ K (Fig. 1d) which is among the highest $T_c$ value ever reported for CoSi$_2$ (Ref. [26]).

We first discuss the differential conductance spectra $dI(V,T)/dV$ (where $V$ is the bias voltage, and $I$ is the current) of highly transparent CoSi$_2$/TiSi$_2$ S/N tunnel junctions like the one depicted in the optical micrograph in Fig. 1a. Figure 1b shows cross-sectional TEM images of the junction cut by a focused ion beam along the dashed line indicated in Fig. 1a. In the right panel, an enlarged version of the interface region indicated by the red arrow in the left panel, is presented. It reveals a well-defined atomic arrangement at the interface. The corresponding diffraction patterns of CoSi$_2$ and TiSi$_2$ are also shown. Figure 1c schematically depicts the junction structure and our $dI/dV$ measurement configuration. Below $T_c$, the total resistance is given by $R(T) = R_J(T) + R_{N2}$, where $R_J$ denotes the junction resistance of the S/N interface, and $R_{N2}$ the residual resistance of the TiSi$_2$ counter-electrode which can be independently determined (S4) and subtracted from $R(T)$.

The S/N interface of our junctions is right-angle shaped, see Fig. 1c. The top interface, indicated by a blue-dash line, is opaque due to the fabrication process. The clean vertical interface immersed in and protected by the Si(100) substrate, indicated by a red line, dominates the measured Andreev tunneling spectra (S5). Figure 1d shows the zero-bias junction conductance $G_J(V=0,T) = 1/R_J(T)$ for a typical device, referred to as J1, below 2 K. An increase of $G_J(0,T)$ with decreasing $T$ for $T < 1.48$ K is evident. A second conductance increase setting in at ~ 0.65 K is also clearly discernable. The excess tunneling conductance (blue curve) is fully suppressed in $B = 0.12$ T (red curve). The small conductance decrease forming a dip at 1.48 K is caused by the contribution from the dirty top interface (S5 and S6). This small drop does not affect the superconducting $dI/dV$ characteristics and will be ignored below. The inset shows the resistivities of the $CoSi_2$ and $TiSi_2$ films below 2 K.

Figure 2a shows the $dI/dV$ curves for device J1 between 0.37 and 1.78 K in $B = 0$. At 0.37 K, the $dI/dV$ curve displays energy gap signatures (i.e., two symmetric side dips[19,20]) at bias voltages $V_g \approx \pm 0.225$ mV and a broad conductance hump spreading across an $eV$ range equivalent to the energy gap, where $e$ is the electronic charge. This broad hump arises from Andreev bound states with linear dispersion[22]. It gradually diminishes as $T$ is increased to $T_c$. As shown in (S7), the extracted $T$-dependence of the gap amplitude $\Delta(T) = |eV_g(T)|$ deviates from the weak-coupling Bardeen-Cooper-Schrieffer (BCS) expression. Regarding the hump feature, we recall that the $dI/dV$ spectra for the $d$-wave superconductor $YBa_2Cu_3O_{7-\delta}$ possess a sharp peak at zero bias, instead of a broad hump, due to its Andreev bound states having a flat dispersion[25,26]. Figure 2a also shows that the $dI/dV$ spectra feature a second, narrow ZBCP that is visible for $T \leq 0.65$ K. This ZBCP is responsible for the increase of $G_J(V = 0, T \leq 0.65$ K$)$ shown in Fig. 1d. We will comment on its origin below.

Figure 2b shows the $dI/dV$ curves of device J1 for several $B$-field values at $T = 0.37$ K. It demonstrates that the two-step ZBCP is progressively suppressed as $B$ is increased, with the narrow peak first vanishing at $B^* \approx 60$ mT. The two-step ZBCP cannot be due to thermal effects as the accompanying spikes at high bias $V > V_g$ (Figs. 2a and b) signal a current density equal to the local critical current density of $CoSi_2$. Figures 2c to e show the normalized $(dI/dV)_n$ curves for three different junctions (devices J1, J2, J3) at $T = 0.37$ K and in $B = 0$, where $(dI/dV)_n$ denotes the differential conductance normalized to its corresponding normal-state value. Pronounced two-step conductance peaks are seen in every junction, though the details of the lineshapes vary from junction to junction.

Interestingly, these lineshapes neither fit expectations based on pure $s$-wave nor $d$-wave pairing potentials[18] (S8). We thus have analyzed the spectra by applying a chiral $p$-wave gap function[12,24], assuming that the $CoSi_2/TiSi_2$ interface lies in the $y$-$z$ plane. Notably, the differences in lineshape can be (almost completely) ascribed to the varying magnitudes of the dimensionless tunneling barrier height $Z$ which cannot be fully controlled experimentally. The broad hump is essentially flat in Fig. 2e, implying $Z \ll 1$ in this particular junction. The red solid curves in Figs. 2c to e are fits to the data based on a chiral $p$-wave gap function[11,24] with the fitted values $Z = 0.81$ (c), 0.67 (d) and 0.10 (e) (S8). Since the S/N junctions are made of diffusive (TiSi$_2$) thin films which possess higher than average effective tunneling transparencies, correspondingly smaller Z values are expected, in line with the fitted values[27].

The consistently high quality of these fits points to a dominating *p*-wave pairing amplitude in CoSi$_2$/TiSi$_2$ interfaces. These fits which are based on the clean limit of TiSi$_2$ capture very well the overall lineshape of the $dI/dV$ spectra but fail to reproduce the sharp, second ZBCP (this second ZBCP will be attributed to the proximity effect in *diffusive* chiral *p*-wave S/N junctions[28], see below). We note that the *T*, *V*, and *B* dependence shown in Figs. 1d, 2a, and 2b rule out magnetic Kondo impurities as a possible origin of this ZBCP. Another possible source of the central ZBCP could be reflectionless tunneling due to the presence of disorder in TiSi$_2$. We can also rule out this phase-coherent Andreev scattering[29] as the underlying reason of the ZBCP due to its *V* and *B* dependences. The threshold magnetic field of the ZBCP is much larger than that expected for phase-coherent Andreev reflection which can be estimated via $B_{\varphi,c} = \hbar/(eL_\varphi t_N)$, where $\hbar$ denotes the Planck constant, $L_\varphi$ the phase-coherent length of TiSi$_2$ ($\approx 1$ μm at 0.37 K, S9), and $t_N$ the thickness of TiSi$_2$ ($\approx 125$ nm, Methods). This results in an expected $B_{\varphi,c}$ for reflectionless tunneling of the order of a few mT.

The diffusion constant of TiSi$_2$ is estimated as $D_{DN} \approx 65$ cm$^2$/s while the FWHM ($\approx 0.04$ meV) of the narrow ZBCP is of the same order of magnitude as that observed in the T-shaped proximity structures discussed below (cf. Fig. 3c, S10, and Table S2). This FWHM reflects the Thouless energy of the diffusive normal metal (DN) $E_{Th} = \hbar D_{DN}/L_{DN}^2$, where $L_{DN}$ is the characteristic length of a diffusive segment of TiSi$_2$ bound by grain boundaries (cf. Fig. 1c and S2). Thus, a $L_{DN}$ can be estimated from $E_{Th}$ and $D_{DN}$, whose value $\approx 330$ nm is in line with the $L_{DN}$ evaluated from scanning electron microscopy (SEM) and TEM images (S2). This points to the narrow ZBCP as being intrinsic to diffusive CoSi$_2$ junctions. This is further corroborated by the magnitude of the threshold magnetic field (cf. the threshold *B* field in the T-shaped proximity structures discussed below). Indeed, since $E_{Th} \ll \Delta$, such a narrow ZBCP has been predicted to occur in diffusive chiral *p*-wave S/N junctions[28].

As an independent and complementary analysis to further substantiate this finding, we analyze conductance spectra of highly transparent CoSi$_2$/TiSi$_2$ proximity structures[23]. Figure 3a shows a schematic diagram of such a T-shaped proximity structure. It was shown by Asano et al.[12] that the Cooper pairs from a spin-triplet S penetrating into N will be transferred into *odd*-frequency spin-*triplet even*-parity *s*-wave pairs which lead to an enhanced quasiparticle local density of states (LDOS) at the Fermi energy $E_F$ of N[12,33]. In contrast, for any spin-singlet S, a zero-bias dip in $(dI/dV)_n$ is expected due to the reduction of LDOS at $E_F$ in N[12,33]. This is illustrated in Fig. 3a and explains the suppression of the LDOS at $E_F$ for such a proximity structure of superconducting aluminum mentioned above[24,30]. The inset of Fig. 3b shows a SEM image (false-colored for clarity) of our T-shaped proximity structure, referred to as device A1. The measured $(dI/dV)_n$ curve of device A1 exhibits a characteristic ZBCP as shown in Fig. 3c which confirms the existence of a spin-triplet pairing amplitude in S[12,33].

At $T = 0.37$ K, the FWHM of the ZBCP is $\approx 0.03$ meV, see also Fig. 3d. This is consistent with the independently determined Thouless energy $E_{Th}$ ($\approx 0.05$ meV, S10 and Table S2) of the penetrated Cooper pairs undergoing diffusive motion in TiSi$_2$. Figure 3d demonstrates that the ZBCP (at 0.37 K) of the same device is gradually suppressed by an increasing $B$ field, and completely vanishes at $B^* \approx 50$ mT, a value that is smaller than the critical magnetic field of CoSi$_2$ ($B_{c//} \approx 100$ mT, cf. S3). The value of $B^*$ is comparable to the threshold value at which the sharp ZBCP observed in CoSi$_2$/TiSi$_2$ S/N junction vanishes (cf. Fig. 2b) and thus points to a common origin.

The inset of Fig. 3e shows a false-color SEM image of a further T-shaped device, referred to as A4, which is composed of a more resistive form of TiSi$_2$ than device A1 (S10 and Table S2). The main panel shows the pronounced ZBCPs occurring below $T \approx 1.35$ K. In Fig. 3f, the ZBCP is suppressed by an increasing $B$ field and vanishes at $B^* \approx 100$ mT at $T = 0.37$ K. The enhanced ZBCP occurring in the more resistive DN fits well the theoretical expectation of a triplet-superconductivity induced proximity effect[23]. Thus, the conductance spectra of T-shaped CoSi$_2$/TiSi$_2$ structures indicate the presence of triplet superconductivity. To the extent that conductance spectroscopy can distinguish between helical and chiral triplet superconductivity, these spectra are compatible with chiral $p$-wave superconductivity.

The presence of chiral $p$-wave superconductivity in CoSi$_2$/TiSi$_2$ heterojunctions implies the existence of supercurrents associated with chiral domains. In order to test for their existence we turn to a careful analysis of the magnetoresistance of the S/N junctions[22,31]. Intriguingly, the magnetoresistance displays a peculiar hysteretic behavior. Figure 4a shows the zero-bias magnetoresistance curves for device J4 measured in sweeping in-plane $B$ fields and at three different $T$ values. The arrows and numbers indicate the $B$ sweeping sequence. At 1.74 K, no magnetoresistance is detected. Once $T$ decreases below $T_c$, a hysteresis appears. The size of the hysteresis, indicated by $\Delta B/2$, is ~ 5.5 mT (S11). A hysteresis of $\Delta B/2 \sim 20$ mT is found in device J5 at 0.37 K as can be read off from Fig. 4b. Figure 4c shows the magnetoresistance curves for device J2. Note that in every junction the magnetoresistance minimum occurs before $B$ passes through zero. This 'advanced' feature is reminiscent of the anomalous magnetic response reported for Bi/Ni bilayer superconductor[32] and $Sr_2RuO_4$ (Ref. [33]), and theoretically interpreted by Bouhon and Sigrist[34]. It is, however, incompatible with the hysteresis expected for magnetic flux pinning in a conventional superconductor.

Microscopically, in a chiral superconductor, internal magnetic fields will be induced by spontaneous supercurrents which flow along the domain surfaces[35,36]. In a sample containing several chiral domains, these magnetic fields will not fully average out and hysteresis can appear. Neighboring domains may possess opposite chiralities ($p_x \pm ip_y$) and corresponding orbital angular momentum of a Cooper pair described by $L_z = \pm\hbar$. As a characteristic average size of these chiral domains, we take the mean grain size, which in our $CoSi_2$/Si(100) films have an average lateral size of ~ 300 nm (Ref. [25]). Thus, for our junctions with typical areas ~ (0.05 – 1) μm² (Table S1), the Andreev tunneling currents will flow through only a limited number (~ 10) of domains. Consequently, a net local field persists, causing a hysteretic magnetoresistance. In contrast, devices J1 and J3 have comparatively large S/N interface areas, allowing for an averaging out of different chiralities. As a result, a considerably diminished hysteresis in the junction magnetoresistance is expected in larger junctions. This is indeed what we find (S11). This should be compared with a vortex-pinning driven hysteresis, where all vortices should have the same helicity in a non-zero magnetic field so that no averaging out occurs. As mentioned above, $\kappa$ is only slightly above $1/\sqrt{2}$, which not only implies a $B_{c1}$ close to $B_{c2}$, but also that the flux lines in the vortex phase only weakly interact. Thus, no strong hysteresis is expected from vortex pinning[37].

Specifically, as the lower critical field in our films can be estimated to have a lower bound of 62 mT (S3), no vortex contribution is expected at the field strengths where the hysteresis minimum occurs. For large positive (negative) fields, the magnetoresistance slope has to be positive (negative) as the normal-state behavior is approached, while the field-derivative near zero field in a down-sweep ($|B| \to 0$, so that one chirality dominates) is negative for positive chirality and vice versa[22,31]. Thus, in a down-sweep before reaching $B = 0$ a minimum in the magnetoresistance has to occur. This unique hysteretic behavior thus provides yet another indication for chiral superconductivity in $CoSi_2/TiSi_2$ heterojunctions[32,33]. The winding number of the vortex in the domain of a chiral $p$-wave state is either 0 or +2, depending on the $B$ field direction and the chirality[5,33], and may lead to step-like changes in magnetoresistance. At present, we are unable to resolve such tiny changes in the magnetic response. This may be related to the comparatively large lower critical field $B_{c1}$ related to the comparatively small value of the Landau-Ginzburg $\kappa$ parameter of $CoSi_2$.

Taken the magnetoresistance hysteresis together with the results for the S/N junctions and the T-shaped proximity structures, our observations indicate that $CoSi_2/TiSi_2$ heterostructures feature a dominant triplet pairing amplitude whose order parameter appears to be compatible with chiral $p$-wave pairing. Interestingly, we find that $CoSi_2$ interfaces feature an unexpectedly large spin-orbit coupling[38] with a characteristic energy scale ($\Delta_{so} \approx 6$ meV, S9) corresponding to $\sim 30\Delta$, where $2\Delta$ is the superconducting gap. It is known that time-reversal invariant odd-parity pairing arises naturally in such a situation[12]. While it may be difficult to distinguish helical from chiral $p$-wave superconductivity based solely on Andreev spectroscopy, the unusual features of the magnetotransport point to the presence of chiral Cooper pairing. In this context, we recall that the observed deviation from the BCS gap equation mentioned above might result from the different $3d$ electron bands and their expected bandwidth narrowing in the vicinity of the interface. On the other hand, future thermodynamic and zero-field muon spin relaxation and optical Kerr measurements will be useful in discriminating if the triplet superconductivity is a bulk property of $CoSi_2$. As the $CoSi_2$ interfaces possess a comparatively high quality[25], our findings should also prove helpful in relation to addressing other systems with possible chiral $p$-wave pairing like $Sr_2RuO_4$ (Refs. 10,33). In particular, given that the fabrication processes and scalability of epitaxial $CoSi_2/Si$ heterostructures are fully compatible with state-of-the-art silicon-based integrated-circuit technology[39], we expect that our observation can provide a viable route for realizing novel $p$-wave-based devices in silicide-based material system.

**Acknowledgments:** We thank S. K. Yip, B. Rosenstein, C. D. Hu, V. Mishra, Y. Li, J. Y. Lin and J. D. Thompson for discussions. **Funding:** This work was supported by the Taiwan Ministry of Science and Technology through grant numbers MOST 106-2112-M-009-007-MY4 and 108-2811-M-009-500, and by the Center for Emergent Functional Matter Science of National Chiao Tung University from the Featured Areas Research Center Program within the framework of the Higher Education Sprout Project by the Ministry of Education (MOE) in Taiwan. Work at Zhejiang University was in part supported by the National Key R&D Program of the MOST of China, grant number 2016YFA0300200 and the National Science Foundation of China, grant number 11774307. **Author contributions:** S.P.C., C.C.T. and J.J.L. conceived the experiment. S.P.C. and S.S.Y. carried out sample fabrication and transport measurements. All authors participated in the discussion, analysis and interpretation of data. S.P.C., S.K. and J.J.L. wrote the manuscript. **Competing interests:** Authors declare no competing interests.

**Data and materials availability:** All data is available in the main text or the supplementary materials.

**Methods**

**Junction fabrication.** We used two-step thermal-evaporation deposition and thermal annealing to fabricate $CoSi_2$/$TiSi_2$ junctions (see S1 for schematic fabrication procedures). In the first step, a 30-nm-thick Co film was deposited on an electron-beam-lithographically (EBL) defined micrometer-wide area of an undoped Si(100) substrate. The deposited Co film on Si(100) substrate was annealed at 700°C for 1 h and then at 800°C for 1 h to form a 105-nm-thick $CoSi_2$ film[15]. Because $CoSi_2$ is the final high-temperature phase along the phase formation sequence ($Co_2Si \rightarrow CoSi \rightarrow CoSi_2$) with rising annealing temperature, under this Si-rich condition and with a long period of annealing time, all Co atoms were transformed into a single-phased, paramagnetic $CoSi_2$ (Ref. [40]). In this step, Co atoms were the major moving species, namely, they diffused downward into the Si(100) substrate to form an epitaxial $CoSi_2$/Si(100) heterostructure[39]. After formation of the compound $CoSi_2$, Co atoms are connected to Si atoms in an eightfold coordinated structure through strong covalent bonding and therefore Co atoms lose their diffusivity[39,40]. The $CoSi_2$/Si(100) heterostructure had high thermal stability up to 900°C and a melting point of 1326°C (Ref. [39]).

After annealing, the Si(100) substrate with $CoSi_2$/Si(100) heterostructure was exposed to air for the second step of the EBL process. A 50-nm-thick Ti film was deposited on the patterned area close to the $CoSi_2$/Si(100) heterostructure. The deposited Ti was annealed at a temperature between 720 and 800°C to form a 125-nm-thick $TiSi_2$ film. Different annealing temperatures resulted in different device properties and thus tuned the device parameters. During this annealing process, Si atoms were the major moving species[39], i.e., the Ti atoms had negligible inter-diffusion with the compound $CoSi_2$. This fabrication method led to the formation of a high-quality "active" $CoSi_2$/$TiSi_2$ junction (indicated by a vertical red line in Fig. 1c and Fig. S1D) which was immersed in the Si(100) substrate and protected by a "passivated" $TiSi_{2-x}$ thin layer (see Figs. 1b and c).

We had patterned the EBL area so that the $TiSi_2$ counter-electrode was in contact and overlapped the $CoSi_2$/Si(100) heterostructure (see Fig. S1). This protruding $TiSi_2$ section of a few μm long was an indispensable design. It served as a passivating layer to protect the active $CoSi_2$/$TiSi_2$ interface from oxidation and contamination during the second-step annealing process. On the other hand, because the top surface of the $CoSi_2$/Si(100) heterostructure was exposed to air and was dirty, this passivated $CoSi_2$/$TiSi_2$ interface (indicated by a horizontal blue dashed line in Fig. S1D) contributed only a negligible Andreev tunneling current at $T < T_c$. Our control $dI/dV$ measurements indicated that the junction resistances of the top passivated

CoSi$_2$/TiSi$_2$ interfaces were two to three orders of magnitude larger than those of the active CoSi$_2$/TiSi$_2$ junctions (S5).

**Junction structure characterizations.** Several junctions were processed by using the focused ion beam technique (TESCAN GAIA3) to make specimens for the CoSi$_2$/TiSi$_2$ interface structure studies. The structure characterizations were carried out on a field emission transmission electron microscope (JEOL JEM-F200).

**Electrical measurements.** Ti/Au (15/65 nm) bonding pads were deposited on the CoSi$_2$/TiSi$_2$ (S/N) tunnel junction and T-shaped proximity structure devices via electron-gun evaporation through a mechanical mask. Thin Au wires were attached to the bonding pads by a wire bonder. Devices were mounted on the sample holder of an Oxford Heliox $^3$He cryostat equipped with a 2-T superconducting magnet or in a BlueFors LD-400 dilution refrigerator equipped with a 9-T superconducting magnet. The $dI/dV$ and magnetoresistance curves were measured using the four-probe method. The applied current $I$ was comprised of a dc component ($I_{dc}$) and an ac component ($I_{ac}$). The two components were added by a homemade bias circuit. The $I_{dc}$ was generated by a Keithley Model 6221 dc current source, while the $I_{ac}$ was generated by a Linear Research Model LR700 ac resistance bridge operating at 16 Hz. The dc voltage drop across the device was measured by a Keithley Model 2182 nanovoltmeter. The differential resistances (conductances) were registered by the LR700 ac resistance bridge.

Figure 1

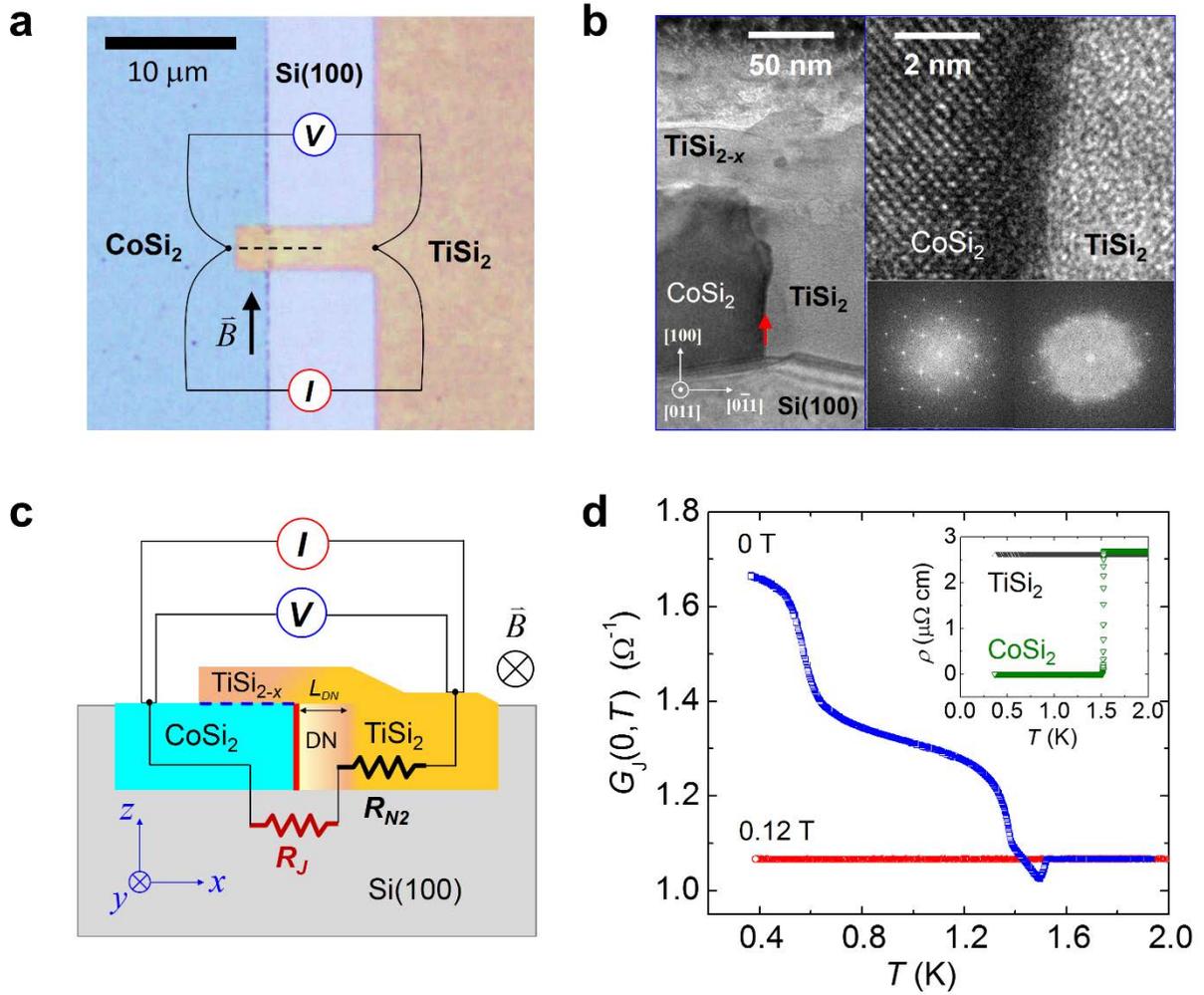

**Fig. 1. CoSi$_2$/TiSi$_2$ tunnel junctions and conductance spectra. a**, An optical micrograph (top view) of a CoSi$_2$/TiSi$_2$ tunnel junction on a Si(100) substrate, with a schematic *I-V* characteristic measurement configuration superimposed. **b**, Left: cross-sectional TEM image of a CoSi$_2$/TiSi$_2$ junction. Right: a zoom-in of the interfacial region indicated by the red arrow in the left panel. Below: diffraction patterns of CoSi$_2$ and TiSi$_2$. **c**, Schematic diagram depicting junction structure and device resistance components. $R_J$ denotes the junction resistance, and $R_{N2}$ denotes the residual resistance of the TiSi$_2$ counter-electrode. The vertical red line indicates the active CoSi$_2$/TiSi$_2$ junction. The horizontal blue dashed line is a high-resistance CoSi$_2$/TiSi$_{2-x}$ interface. $L_{DN}$ denotes the length scale of a diffusive normal metal (DN). **d**, Main panel: $G_J(V=0,T)$ curves below 2 K for device J1 in $B=0$ and 0.12 T. The inset shows the resistivities of the constituent CoSi$_2$ and TiSi$_2$ films below 2 K.

Figure 2

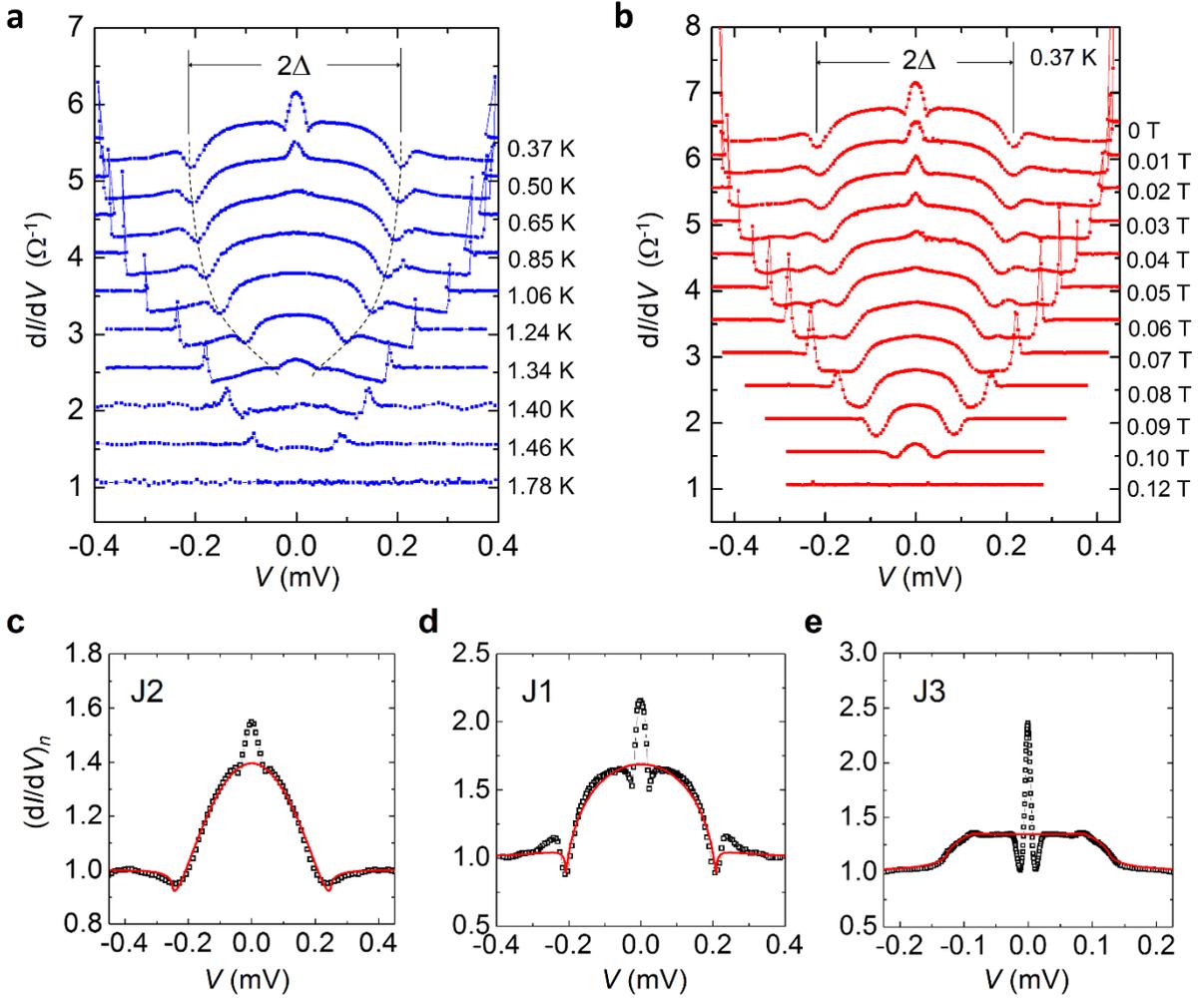

**Fig. 2. Conductance spectra of CoSi$_2$/TiSi$_2$ tunnel junctions. a**, $dI/dV$ curves for device J1 at several $T$ values, as indicated, and in $B=0$. The dashed curves indicate the superconducting gap $2\Delta$. **b**, $dI/dV$ curves for the same device in several $B$ values, as indicated, and at $T=0.37$ K. In **a** and **b**, the $dI/dV$ curves are vertically offset for clarity. The spikes outside the superconducting gap occur when the applied bias current matches the superconducting critical current of the device. **c** to **e**, Normalized $(dI/dV)_n$ curves for three devices at $T=0.37$ K and in $B=0$. These tunnel junctions were fabricated under nominally similar conditions. The junction resistances are $R_J(0.37\text{ K})=0.92$ (**c**), 0.61 (**d**) and 0.16 (**e**) $\Omega$. The red solid curves in **c** to **e** are theoretical predictions for a chiral $p$-wave pairing superconductor with fitted barrier parameter $Z=0.81$ (**c**), 0.67 (**d**) and 0.10 (**e**) (see text and S8).

Figure 3

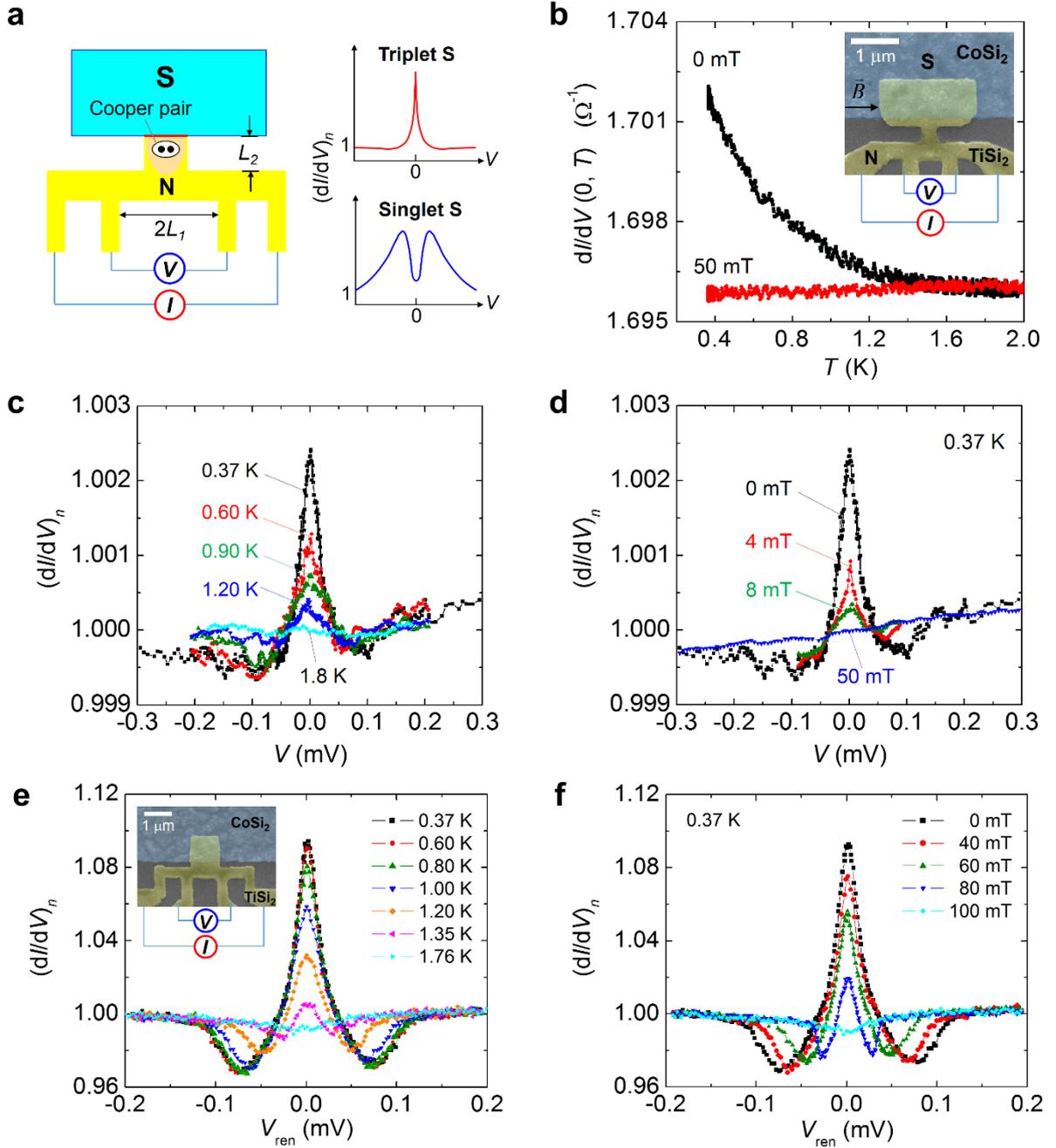

**Fig. 3. T-shaped CoSi$_2$/TiSi$_2$ superconducting proximity structure and conductance spectra. a**, Schematic of a T-shaped superconductor (S)/normal metal (N) proximity structure, with *I-V* characteristic measurement configuration. The sketch also depicts the predicted zero-bias conductance peak (dip) for a spin-triplet (spin-singlet) S. **b**, Inset: a false-colored SEM image of the T-shaped device A1. The scale bar is 1 μm. Main panel shows the zero-bias $dI(0,T)/dV$ curve in $B = 0$ and 50 mT. The proximity effect sets in at $T_{onset} \approx 1.4$ K. **c**, Normalized conductance spectra $(dI/dV)_n$ of device A1 at several *T* values show a ZBCP at $T < T_{onset}$. **d**, At 0.37 K, the ZBCP is gradually suppressed with increasing *B* and completely vanishes at $B = 50$ mT. **e**, Normalized conductance spectra $(dI/dV)_n$ of T-shaped device A4

at different $T$ values. The renormalized bias voltage $V_{ren}$ of the independent axis is defined as the voltage drop over the TiSi$_2$ grain which is responsible for the proximity effect (S10). Inset: a false-colored SEM image of device A4. **f**, At $T = 0.37$ K, the ZBCP of device A4 is gradually suppressed with increasing $B$ and vanishes completely at $B = 100$ mT.

Figure 4

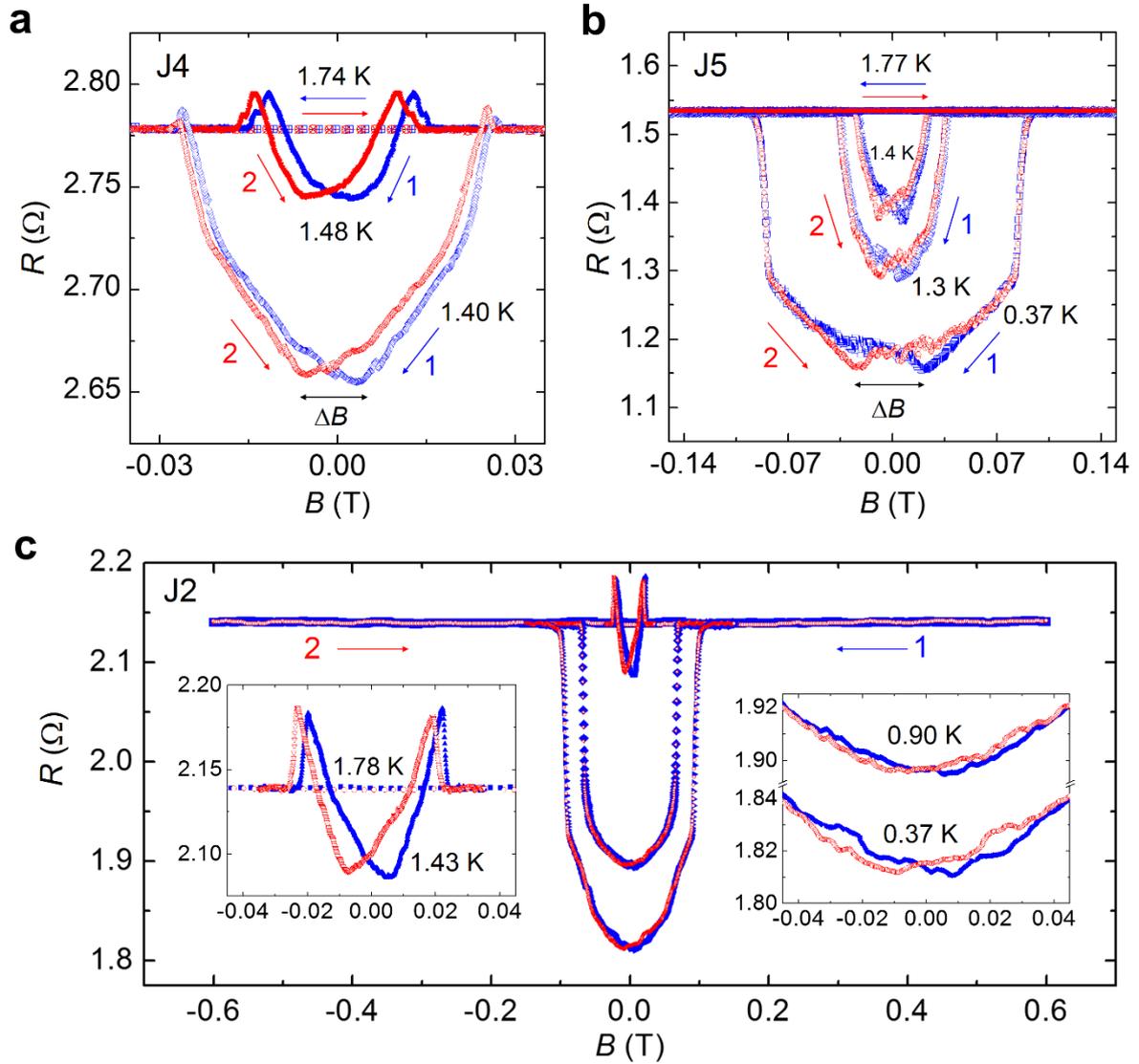

**Fig. 4. Hysteretic zero-bias magnetoresistance in CoSi$_2$/TiSi$_2$ tunnel junctions.** Zero-bias magnetoresistance curves of devices J4 (**a**), J5 (**b**) and J2 (**c**) at several $T$ values, as indicated. In **c**, left (right) inset shows a zoom-in for the magnetoresistance at 1.43 and 1.78 K (0.37 and 0.90 K). The arrows and numbers indicate the sweeping sequence of the magnetic field. $\Delta B$ denotes the size of the hysteresis.

# Supplementary Information

## Observation of triplet superconductivity in $CoSi_2$/$TiSi_2$ heterostructures

S. P. Chiu, C. C. Tsuei, S. S. Yeh, F. C. Zhang, S. Kirchner and J. J. Lin

**Supplementary information and figures**

**Figs. S1 to S11:**

**Tables S1 to S2**

**Full references**

## S1. Two-step fabrication processes of CoSi$_2$/TiSi$_2$ heterostructures on Si(100) substrates

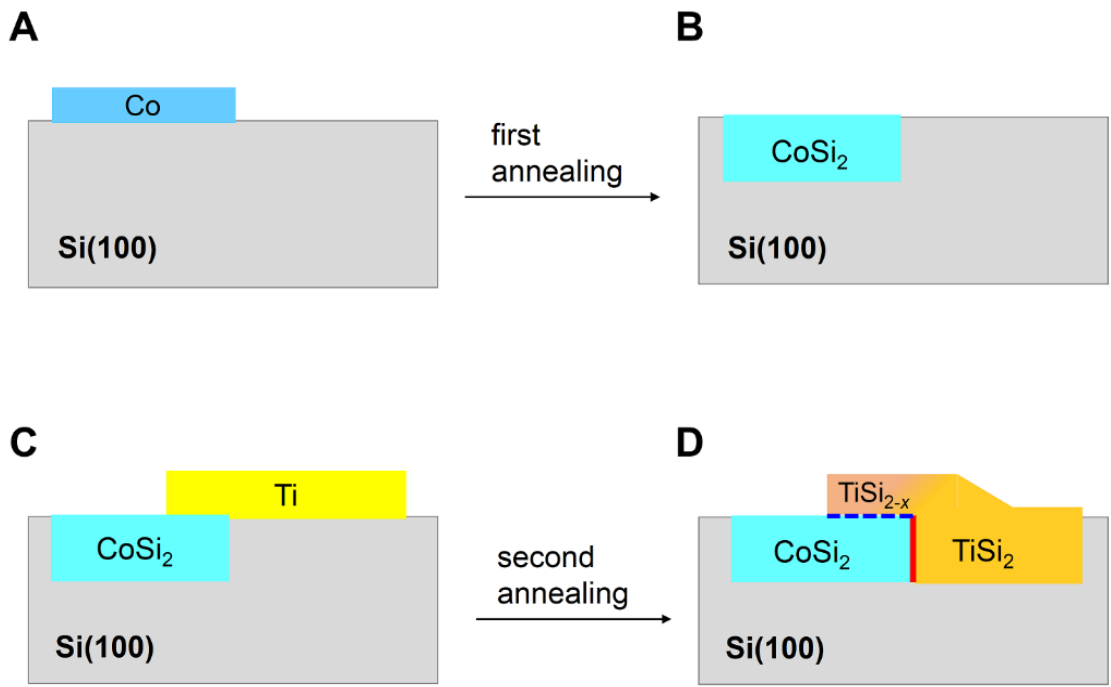

**Figure S1.** Schematic of the two-step fabrication process of CoSi$_2$/TiSi$_2$ tunnel junctions (side view). (**A**) A 30-nm-think Co film was deposited via thermal evaporation on an electron-beam-lithographically (EBL) patterned Si(100) substrate. (**B**) After thermal annealing at 700°C for 1 h and subsequently at 800°C for 1 h, a 105-nm-thick CoSi$_2$ film was formed. (**C**) A 50-nm-thick Ti film was deposited via electron-gun evaporation onto a second EBL patterned area of the Si substrate with the CoSi$_2$/Si(100) heterostructure. (**D**) After thermal annealing at a chosen temperature (between 720 and 800°C for tuning device properties) for a suitable period (e.g., ~1 h), a 125-nm-thick TiSi$_2$ film was grown, which formed a right-angle shaped CoSi$_2$/TiSi$_2$ interface, as indicated by a red solid line and a blue dashed line. The former is referred to as the "active" CoSi$_2$/TiSi$_2$ tunnel junction and the latter as the "passivated" CoSi$_2$/TiSi$_{2-x}$ interface. Between steps (**B**) and (**C**), the as-grown CoSi$_2$/Si(100) heterostructure was exposed to air during the second-step EBL process. The segment of the TiSi$_{2-x}$ thin layer on top of the CoSi$_2$/Si(100) heterostructure is slightly non-stoichiometric, as revealed by energy-dispersive X-ray spectroscopy studies.

## S2. Microstructure of CoSi$_2$/TiSi$_2$ junctions

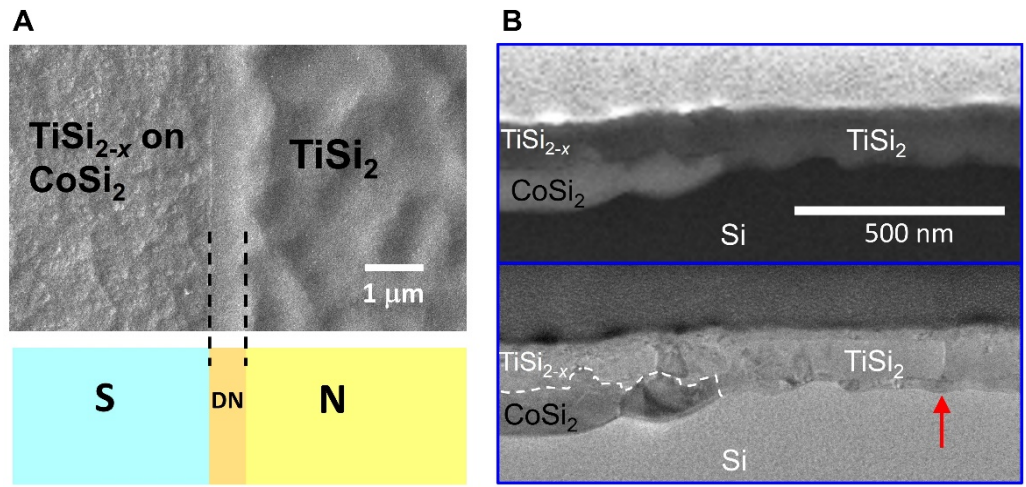

**Figure S2.** (**A**) The upper panel is a top-view scanning electron microscopy (SEM) image of a CoSi$_2$/TiSi$_2$ junction. The TiSi$_2$ part possesses a ripple-like surface profile which gives way to a smooth strip-shaped area formed along the boundary line of CoSi$_2$/TiSi$_2$. The width of this area ranges from 300 to 1000 nm which is related to the distribution of grain boundaries in TiSi$_2$. We interpret this area as the diffusive normal metal part and label it as DN in the schematic below the image. The size of DN can be used to estimate the Thouless energy $E_{Th}$.
(**B**) The upper panel shows a cross-sectional dark-field image of a CoSi$_2$/TiSi$_2$ junction using the scanning transmission electron microscopy (STEM). In this STEM image different materials (CoSi$_2$, TiSi$_2$, and Si) can be well distinguished. The lower panel shows the same area under high-resolution transmission electron microscopy (HR-TEM). The dashed curve indicates the boundary between CoSi$_2$ and TiSi$_2$ (or TiSi$_{2-x}$). A grain boundary of TiSi$_2$ is highlighted by a red arrow which is located about 500 nm away from the CoSi$_2$/TiSi$_2$ interface.

## S3. Critical magnetic field of an epitaxial CoSi$_2$/Si(100) film

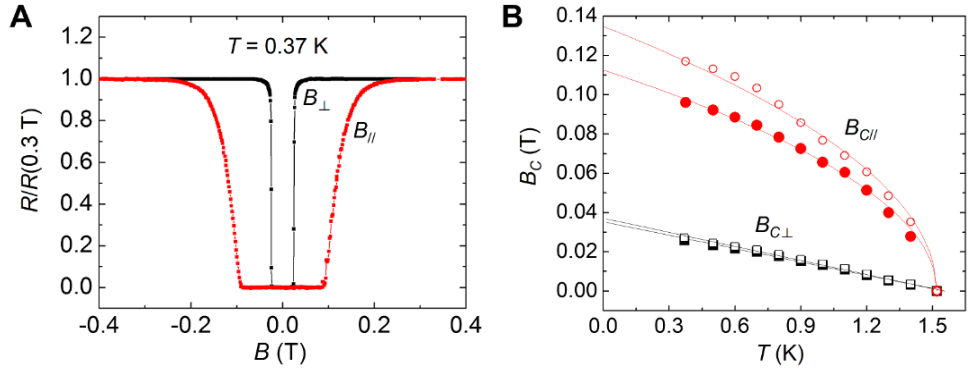

**Figure S3.** (**A**) Normalized resistance $R(B)/R(B=0.3\text{ T})$ as a function of magnetic field at 0.37 K for a 105-nm-thick CoSi$_2$/Si(100) film. The magnetic field is applied either perpendicular or parallel to the film. The critical field can be determined by the field value at which the sample resistance starts to deviate from zero [solid symbols in (**B**)] or where $R(B)/R(B=0.3\text{ T})$ reaches 0.5 [open symbols in (**B**)]. (**B**) The measured perpendicular and parallel critical magnetic fields versus temperature. The red curves indicate that the parallel critical field obeys a square-root-temperature law [$B_{c//}(T) \propto (1-T/T_c)^{1/2}$], while the black straight lines demonstrate that the perpendicular critical field obeys a linear-temperature law [$B_{c\perp}(T) \propto (1-T/T_c)$]. From the extrapolated zero-$T$ values of $B_{c//}(0)$ and $B_{c\perp}(0)$, we have calculated the coherence length to be $\xi(T=0) \simeq 90$ nm and the London penetration depth to be $\lambda_L(T=0) \simeq 74$ nm. Hence, the Ginzburg-Landau parameter $\kappa = \lambda/\xi \approx 0.82$. At 0.37 K a lower bound of $B_{c2//} \approx 95$ mT can be obtained from where the resistance starts to deviate from zero[41], and then $B_{c1//} \approx 62$ mT is estimated from the Gorkov-Brandt theory for $\kappa \approx 0.82$ under the assumption of an isotropic Fermi surface[42].

## S4. Resistance and magnetoresistance of CoSi$_2$/Si(100) and TiSi$_2$/Si(100) films and wires

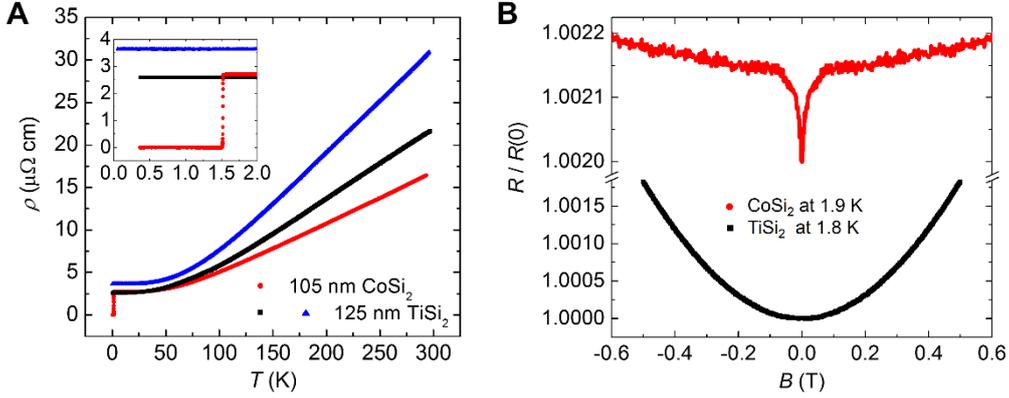

**Figure S4.** (**A**) Dependence of the resistivity on temperature for a 105-nm-thick CoSi$_2$/Si(100) film (red symbols), a 125-nm-thick C54-phase TiSi$_2$/Si(100) film (black symbols) and a TiSi$_2$/Si(100) wire (blue symbols, width $\approx 0.4$ µm), as indicated. The TiSi$_2$/Si(100) film and the electron-beam-lithography patterned wire reveal typical Boltzmann transport behavior down to 0.37 K and 50 mK, respectively, with a large resistance ratio of $R(300\ \text{K})/R(1\ \text{K}) = 8.1 - 8.5$. The inset shows the resistivity for $T < 2$ K. No superconductivity can be found in TiSi$_2$ down to at least 50 mK. The resistivity difference between the TiSi$_2$ film and wire is due to the linewidth effect which emerges at width $< 1$ µm. Flat curves of TiSi$_2$ for $T < 6$ K imply that there are very few magnetic impurities or nonmagnetic defects in C54-phase TiSi$_2$/Si(100). (**B**) Magnetoresistance in a perpendicular magnetic field at 1.8 K for the TiSi$_2$/Si(100) film shown in (**A**). In the low $B$ regime, no sign of a positive magnetoresistance is detected to within our experimental uncertainty, which indicates the absence of discernable weak-antilocalization effect. For comparison, the CoSi$_2$/Si(100) film shown in (**A**) reveals notable positive magnetoresistance in small $B$ fields. [See S9 for a quantitative analysis of the weak-antilocalization magnetoresistance of CoSi$_2$/Si(100) and TiSi$_2$/Si(100) films.]

## S5. Total tunneling conductance of a right-angle shaped CoSi$_2$/TiSi$_2$ interface

The right-angle shaped S/N interface depicted in Fig. S1(D) is composed of an active CoSi$_2$/TiSi$_2$ junction and a passivated CoSi$_2$/TiSi$_{2-x}$ interface. The active junction is highly transparent which is reflected in a small barrier parameter $Z$ ($<1$), and which is almost exclusively responsible for the measured Andreev tunneling spectra at $T < 0.9T_c$. The TiSi$_{2-x}$ thin layer serves only to protect the active junction from oxidation and contamination during the second-step thermal annealing process. From our control experiments, we have found that without this protruding TiSi$_{2-x}$ thin layer, the active junction will always bear a high barrier potential after the second annealing. Consequently, a zero-bias dip rather than a peak was observed in the $dI/dV$ spectra at $T < T_c$. The passivated interface inevitably assumes a large $Z$ value ($\gg 1$), due to the top surface of the CoSi$_2$/Si(100) film being exposed to air and becoming dirty after the remaining processes, as described in S1. Figure S5(A) demonstrates the $dI/dV$ characteristic of such a dirty, top junction, see the caption of Fig. S5.

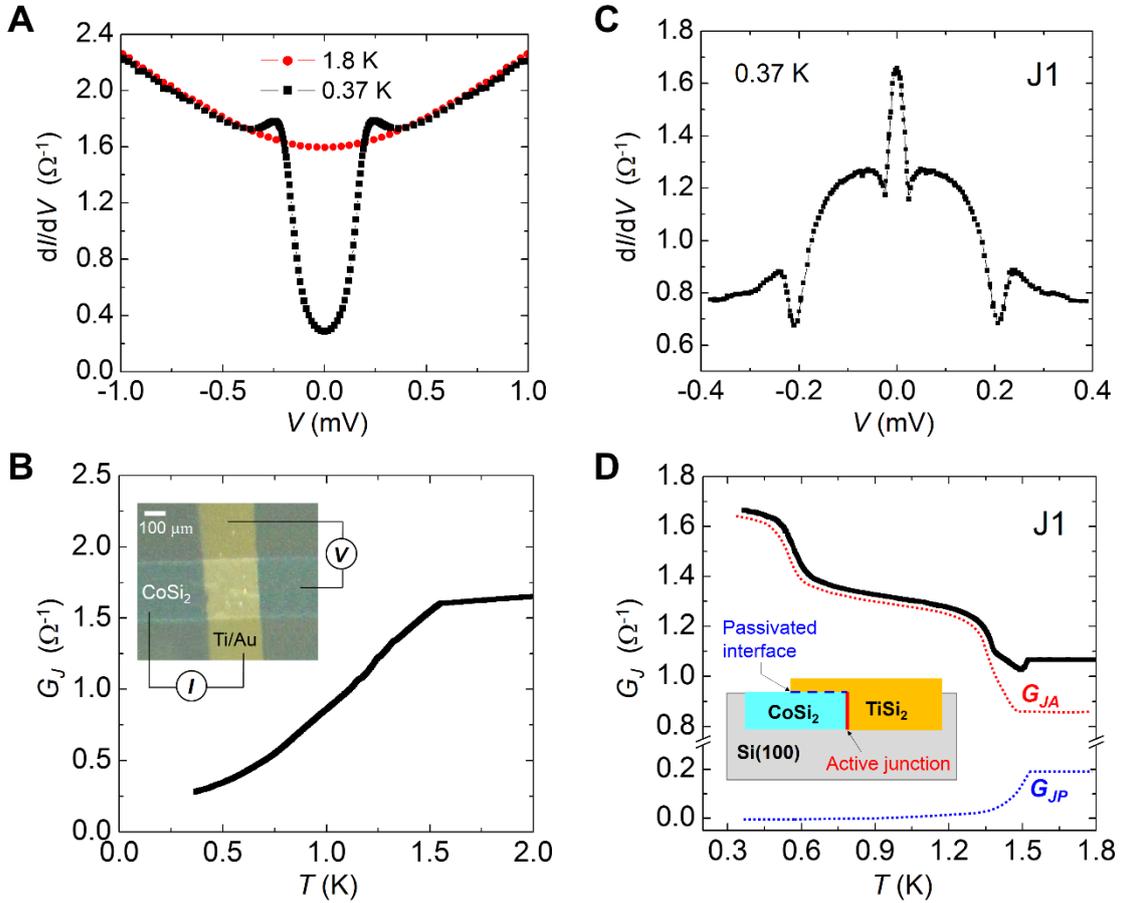

**Figure S5.** (A) $dI/dV$ spectra of a CoSi$_2$/N junction formed on the top surface of a CoSi$_2$/Si(100) film. Here N is a thermally evaporated Ti/Au (15/65 nm) counter-electrode [see the inset in (B)]. A zero-bias dip is seen at 0.37 K. A parabolic background conductance is recovered at 1.8 K. Note that the tunneling conductance per unit area is fairly small ($10^{-5} - 10^{-3}$ $\Omega^{-1}$ μm$^{-2}$), compared to the Andreev tunneling conductance per unit area ($\approx 1$ $\Omega^{-1}$ μm$^{-2}$) through a typical active junction as the one shown in (C). (B) The zero-bias conductance, $G_J = 1/R_J$, as a function of temperature for the same typical junction discussed in (A). A conductance decrease as the temperature decreases below $T_c$ indicates that the CoSi$_2$/N interface is dirty and insulating. Below $T_c$, the junction conductance decreases monotonically with decreasing $T$, indicating that the Andreev reflection process is fully suppressed. The inset shows an optical micrograph of this CoSi$_2$/N junction, with an area of $250 \times 300$ μm$^2$. (**C**) $dI/dV$ curve without normalization at 0.37 K for device J1 shown in Fig. 1d of the main text. (**D**) Zero-bias junction conductance $G_J = 1/R_J$ vs. temperature for device J1. The device-specific conductance dip (resistance peak) in the vicinity of $T_c$ arises from the sum of conductances of the active-junction channel $G_{JA}(T)$ and the passivated-interface channel $G_{JP}(T)$, as schematically (not up to scale) depicted by the red and blue dotted curves, respectively. We note that in the vicinity of $T_c$, both the active junction and the passivated interface contribute to the measured conductance, i.e., $G(T \to T_c^-) = G_{JA}(T \to T_c^-) + G_{JP}(T \to T_c^-)$ . At temperatures well below $T_c$, only the active junction contributes to the Andreev tunneling.

## S6. Junction resistance versus temperature in magnetic fields

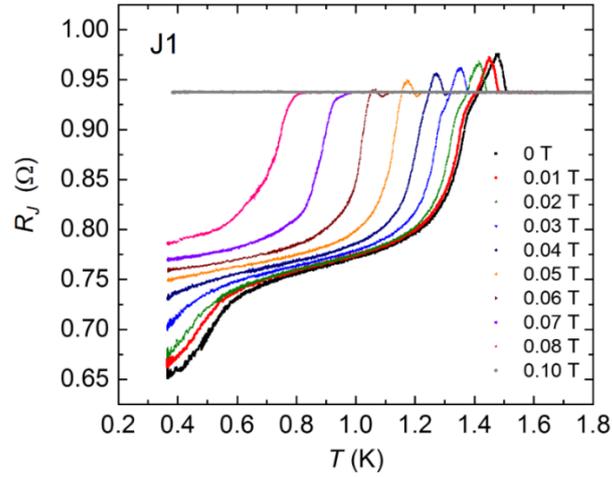

**Figure** S6. Variation of the junction resistance with temperature for junction J1 in magnetic field. The resistance peak in the vicinity of $T_c$ is gradually suppressed and moves to a lower temperature ($T_c$) value with increasing field strength. The 0.60 K (1.48 K) superconducting phase is suppressed by a magnetic field of $\approx 0.05$ ($\approx 0.10$) T. The magnetic field orientation is parallel to the junction plane and to the $CoSi_2/Si(100)$ plane, as indicated in Figs. 1a and c of the main text.

## S7. Superconducting energy gap as a function of reduced temperature

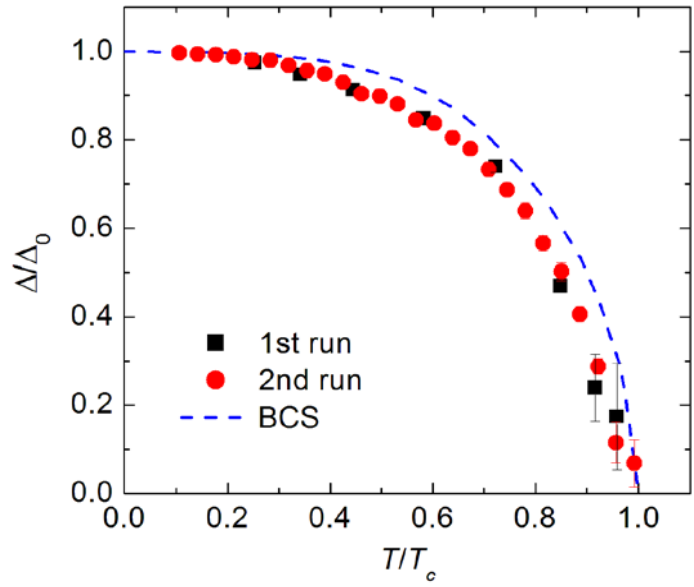

**Figure S7.** Normalized superconducting gap amplitude $\Delta(T)/\Delta_0$ versus reduced temperature $T/T_c$ ($2\Delta_0 = 0.46$ meV and $T_c = 1.46$ K) for device J1 taken from two runs: the first run on a $^3$He cryostat and the second run on a dilution refrigerator. $\Delta(T)$ is determined using the side dips in the $dI/dV$ curves shown in the Fig. 2a of the main text. The measured gap amplitude (symbols) falls systematically and slightly below the weak-coupling BCS theory prediction (solid curve). The side dips at $T/T_c \leq 0.75$ are well developed and the error bars are compatible with the symbol size.

**S8.** Fits of the conductance spectra using predictions for *p*- and *s*-wave superconductors

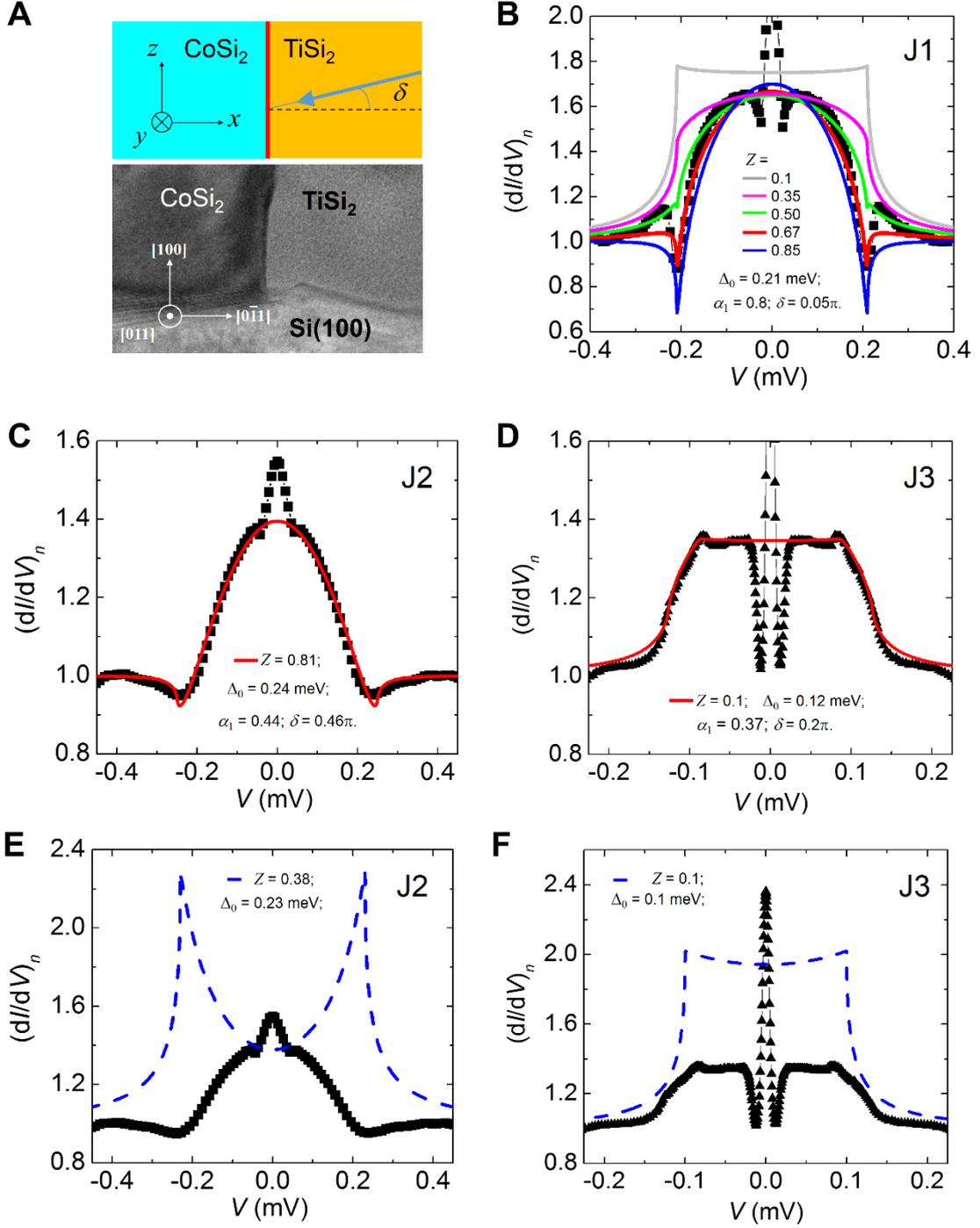

**Figure S8.** For our CoSi$_2$/TiSi$_2$ junctions grown along the Si[100] direction, we consider the incident electrons which predominantly move in the *x-y* plane, within an incident angle $\delta$ indicated in Fig. S8(A). As an approximation, we take $k_z \approx 0$, and thus the $\vec{d}$ vector is reduced to $\vec{d} \approx \hat{z}(k_x + ik_y)$. We use the effective quasi-two-dimensional chiral *p*-wave pair potential

$$\Delta_{\rho\rho'}(\theta,\phi) = \Delta_0 \sin\theta \, e^{i\phi} = \Delta_0 (k_x + ik_y)/|\vec{k}|, \quad (S1)$$

where $\theta$ is the polar angle, $\phi$ is the azimuthal angle in the $x$-$y$ plane, and $\rho$ and $\rho'$ denote spin indices. By using the Andreev reflection coefficient $a_{\rho\rho'}(E,\theta,\phi)$ and the normal reflection coefficient $b_{\rho\rho'}(E,\theta,\phi)$ determined from solving the Bogoliubov–de Gennes equations for the pair potential of Eq. (S1), the normalized tunneling conductance $\sigma(E)$ for an S/N junction in the $y$-$z$ plane and in the clean limit is given by[19,20,43]

$$\sigma(E) = \frac{\int_{\pi/2-\delta/2}^{\pi/2}\int_{-\pi/2}^{\pi/2}[\sigma_{S,\uparrow}+\sigma_{S,\downarrow}]\sigma_N \sin^2\theta\cos\phi\, d\theta d\phi}{\int_{\pi/2-\delta/2}^{\pi/2}\int_{-\pi/2}^{\pi/2}2\sigma_N \sin^2\theta\cos\phi\, d\theta d\phi}, \quad (S2)$$

where $\sigma_{S,\rho}$ and $\sigma_N$ denote the dimensionless superconducting and normal-state tunneling conductances, respectively. Moreover,

$$\sigma_{S,\uparrow} = \frac{1+\sigma_N|\Gamma|^2+(\sigma_N-1)|\Gamma|^4}{\left|1-e^{-2i\phi}(\sigma_N-1)\Gamma^2\right|^2}, \quad \text{with}\ \ \Gamma = \frac{E-\Omega}{|\Delta_0\sin\theta|},\ \ \Omega = \sqrt{E^2-\Delta_0^2\sin^2\theta}, \quad (S3)$$

and

$$\sigma_N = \frac{\sin^2\theta\cos^2\phi}{\sin^2\theta\cos^2\phi+Z^2}, \quad Z = \frac{mH}{\hbar^2 k_F}, \quad (S4)$$

where $E$ denotes the quasiparticle energy measured from the Fermi level. For a unitary superconducting state, one expects $\sigma_{S,\downarrow}=\sigma_{S,\uparrow}$ (Refs. 20,43). In practice, a phenomenological quasiparticle transmission factor $\alpha_1$ ($0<\alpha_1<1$) is introduced to replace $[\sigma_{S,\uparrow}+\sigma_{S,\downarrow}]=2\sigma_{S,\uparrow}$ in Eq. (S2) by $[2\alpha_1\sigma_{S,\uparrow}+2(1-\alpha_1)\sigma_0]$, where $\sigma_{S,\uparrow}(E\gg\Delta_0)\approx\sigma_0=1$ (Ref. 44).

We have performed least-square fits of our $dI/dV$ spectra to the above model, with adjustable junction parameters given by the superconducting energy gap $\Delta_0$, the dimensionless barrier height $Z$, and the angle of the effective incident cone $\delta$, and $\alpha_1$. The fitted values are listed in Figs. S8(B) to (D). Figure S8(B) shows several theoretical curves for different $Z$ values for illustration, which should be compared to the broad hump of device (J1) shown in Fig. 2d in the main text. The best fit is achieved for $Z = 0.67$. Similarly, Figs. S8(C) and (D) show the fitted curves for the broad humps of devices (J2 and J3) shown in Figs. 2c and 2e in the main text, respectively.

For comparison, we show in Figs. S8(E) and (F) fits based on the *s*-wave Blonder-Tinkham-Klapwijk (BTK) model[18] for devices J2 and J3, respectively. The parameters used for the data in Fig. S8(E) are obtained from fitting the zero-bias height of the broad hump of J2. The *s*-wave BTK curve in Fig. S8(F) is created for the same $Z$ value (0.1) as the one extracted in Fig. S8(D) in order to maintain a similar lineshape of the broad hump of J3. We note that only in the extreme case of $Z \to 0$ the lineshapes of theoretical conductance spectra of *s*-wave and chiral *p*-wave S/N junctions are close to each other, but with different zero-bias heights[18,19]. [In the limit $Z \to 0$, the BTK model predicts a plateau of height $(dI/dV)_n = 2$, cf. Fig. S8(F). For a non-zero $Z$ value, the BTK model predicts a split $(dI/dV)_n$ curve, cf. Fig. S8(E).]

## S9. Strong spin-orbit coupling and weak antilocalization effect in epitaxial CoSi$_2$/Si(100) and TiSi$_2$/Si(100) films

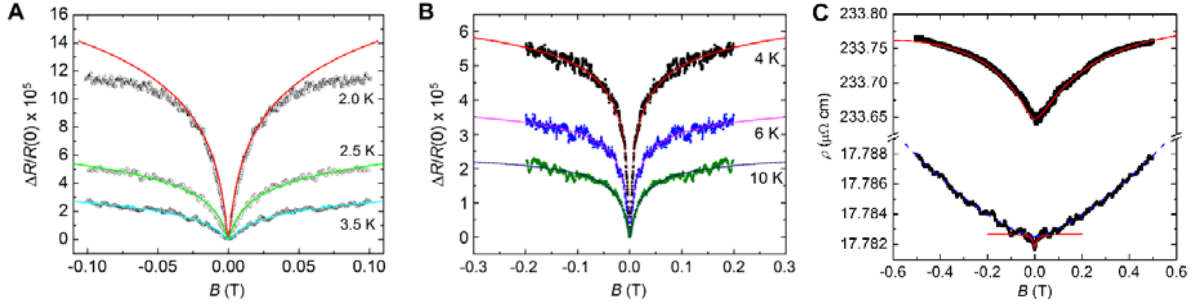

**Figure S9.** (**A**) Magnetoresistance for a 105-nm-thick CoSi$_2$/Si(100) film at three $T$ values above $T_c$. The $B$ field is applied perpendicular to the film. The positive magnetoresistance originates from the weak-antilocalization effect and is indicative of a strong spin-orbit coupling (SOC) in CoSi$_2$. From least-square fits, based on weak-antilocalization theory, the electron dephasing time ($\tau_\varphi$) and the spin-orbit scattering time ($\tau_{so}$) can be extracted. We obtain $\tau_{so} \approx 0.1$ ps, and $\tau_\varphi \approx 110$, 71 and 29 ps and associated electron dephasing lengths $L_\varphi = \sqrt{D\tau_\varphi} \approx 1510$, 1210 and 774 nm at 2.0, 2.5 and 3.5 K, respectively, with the diffusion constant $D = 207$ cm$^2$/s. This spin-orbit scattering time corresponds to a SOC energy $\Delta_{so} = \hbar/\tau_{so} \approx 6$ meV. (**B**) Positive magnetoresistance for a 53-nm-thick CoSi$_2$/Si(100) film at three $T$ values above $T_c$. The extracted scattering times are $\tau_{so} \approx 0.1$ ps, and $\tau_\varphi \approx 41$, 22 and 7.9 ps (corresponding to $L_\varphi \approx 789$, 578, and 347 nm with $D = 152$ cm$^2$/s). The solid curves in (**A**) and (**B**) are the theoretically expected weak-antilocalization corrections which also include the Maki-Thompson superconducting fluctuation effects[45,46]. Note that in both films, $1/\tau_{so} \gg 1/\tau_\varphi$. The strong SOC is an intrinsic property of CoSi$_2$. Our value of $\tau_{so}$ is in agreement with that previously extracted in CoSi$_2$ epitaxial films[38]. (**C**) Magnetoresistance of TiSi$_2$/Si(100) films at 1.9 K and 37.5 nm (upper curve) and 75 nm (lower curve) thickness. The red solid curves are theoretical predictions. The extracted $L_\varphi(1.9\text{ K}) \approx 101$ and 1050 nm, respectively. The associated spin-orbit scattering times are $\tau_{so} \approx 3.6$ (37.5 nm) and 0.16 ps (75nm). The blue dashed curve is the background magnetoresistance due to Lorentz force for the 75-nm-thick film. Our results show that thick TiSi$_2$ films possess much lower residual resistivities and longer dephasing lengths at liquid-helium temperatures than thin films. It is noted that in our CoSi$_2$/TiSi$_2$ T-shaped proximity devices, the TiSi$_2$ part of thickness $\approx 125$ nm possesses even lower residual resistivity which points to a correspondingly longer dephasing length. Note also that, in terms of Boltzmann transport, our TiSi$_2$ films are three dimensional, because the film thickness is larger than the electron elastic mean free path ($t_N \gg l_e \approx 8$ nm).

## S10. Conductance spectra of T-shaped CoSi$_2$/TiSi$_2$ superconducting proximity structures

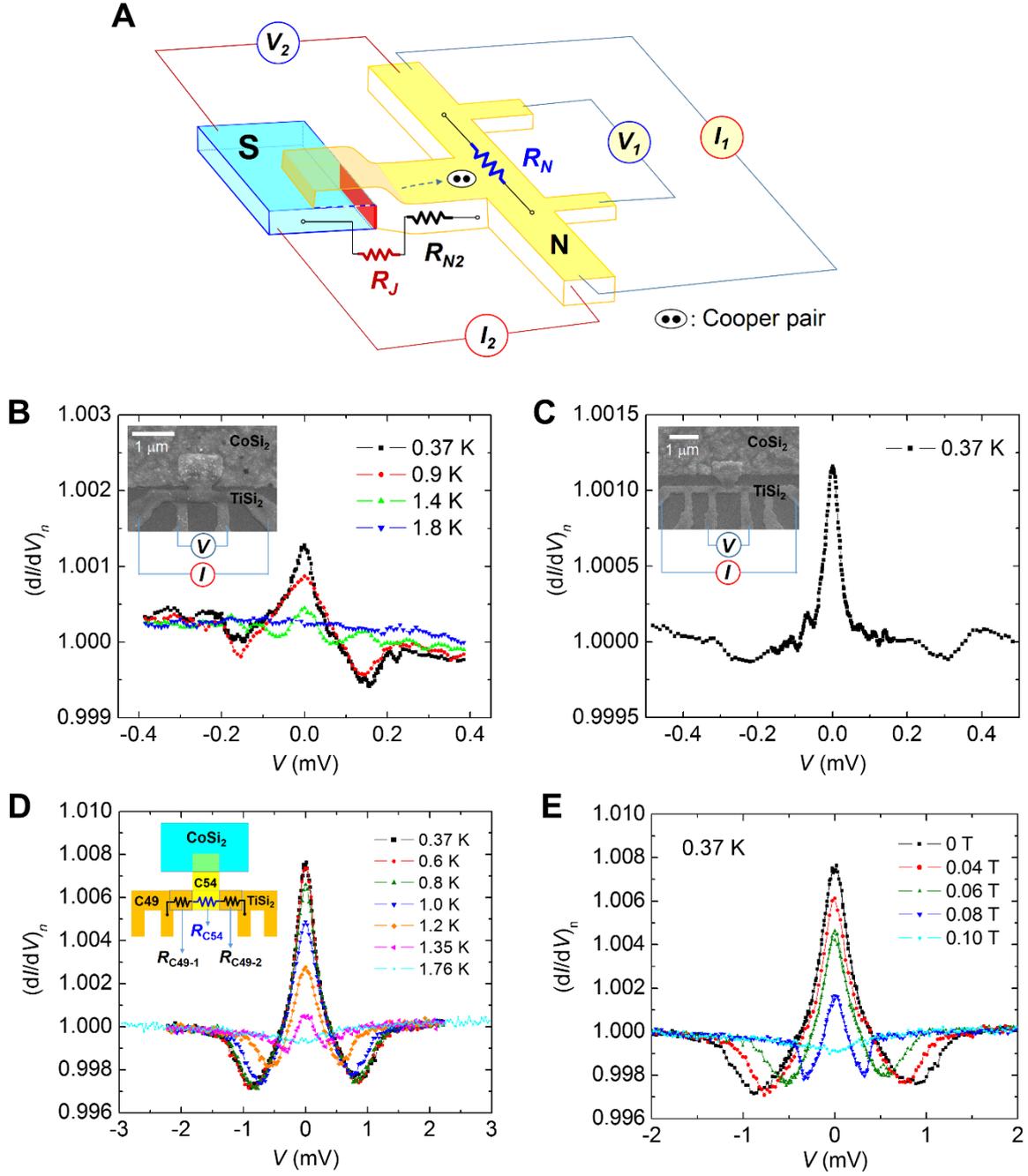

**Figure S10.** T-shaped CoSi$_2$/TiSi$_2$ superconducting proximity structures and normalized conductance spectra $(dI/dV)_n$. (**A**) A schematic of a T-shaped device and the measurement configuration $(I_1, V_1)$ for probing the proximity effect. $R_N$ represents the resistance of the TiSi$_2$ diffusive normal-metal wire. An alternative measurement configuration $(I_2, V_2)$ is used to determine the junction resistance $R_J$ of the CoSi$_2$/TiSi$_2$ tunnel junction. (**B**) to (**D**) show the $(dI/dV)_n$ spectra for three T-shaped proximity devices: A2 (**B**), A3 (**C**), and A4 (**D**). An SEM image of each device is shown in the corresponding inset. (The SEM image of device A4 is shown in the inset of Fig. 3e in the main text.) The scale bar indicates a length of 1 μm. In addition to the ZBCP, two side dips are discernible in each device at $T < T_{onset}$. In the inset of (**D**), a basic circuit model for the resistor $R_N$ of device A4 is shown which is used for estimating the renormalized bias voltage (i.e. the $V_{ren}$ in Figs. 3e and 3f in the main text), as described below. (**E**) At $T = 0.37$ K, the ZBCP in the $(dI/dV)_n$ spectra of device A4 is gradually suppressed with increasing $B$ and completely vanishes at $B = 100$ mT.

There are two distinct metallic phases of TiSi$_2$, referred to as C54-TiSi$_2$ and C49-TiSi$_2$ (Refs. 39,47), which can form in our devices. The C54-TiSi$_2$ phase has a low residual resistivity of $\rho_N < 10$ μΩ cm, and is realized in the T-shaped devices A1–A3 and tunnel junctions J1–J5. The C49-TiSi$_2$ phase is usually formed at a lower annealing temperature[39] and has a relatively high $\rho_N \sim 100$ μΩ cm. The T-shaped proximity device A4 is mainly composed of C49-TiSi$_2$. For this device we have measured the resistance $R_N$ and $R_J + R_{N2}$ by applying the measurement configurations $(I_1, V_1)$ and $(I_2, V_2)$, respectively, shown in Fig. S10(A). The measured values are $R_N = 29.61$ Ω and $R_J + R_{N2} = 1.28$ Ω, indicating that $R_{N2}$ is mainly composed of C54-TiSi$_2$. Therefore, we may construct a simple circuit model for the resistor $R_N$ as illustrated in the inset of Fig. S10(D), where $R_N = R_{C49-1} + R_{C54} + R_{C49-2}$. We also assume that the superconducting proximity effect only extends to the C54-TiSi$_2$ (light-yellow) regime. Thus, the applied bias voltage largely drops along the high-resistivity C49-TiSi$_2$ (dark-yellow) segments. From simulations, we estimate that $R_{C49-1} = R_{C49-2} = 13.5$ Ω and obtain the value of the resistance $R_{C54}$, and thus the relevant voltage drop $V_{ren}$. The measured and renormalized $(dI/dV)_n$ curve is plotted as a function of $V_{ren}$ in Figs. 3e and 3f in the main text. The relevant parameters for the T-shaped proximity devices are listed in Table S2. Note that the FWHMs of the ZBCPs for all devices are of the same order of magnitude as their Thouless energies $E_{Th}$, as predicted[23].

## S11. Absence of a magnetoresistance hysteresis in large $CoSi_2/TiSi_2$ tunnel junctions

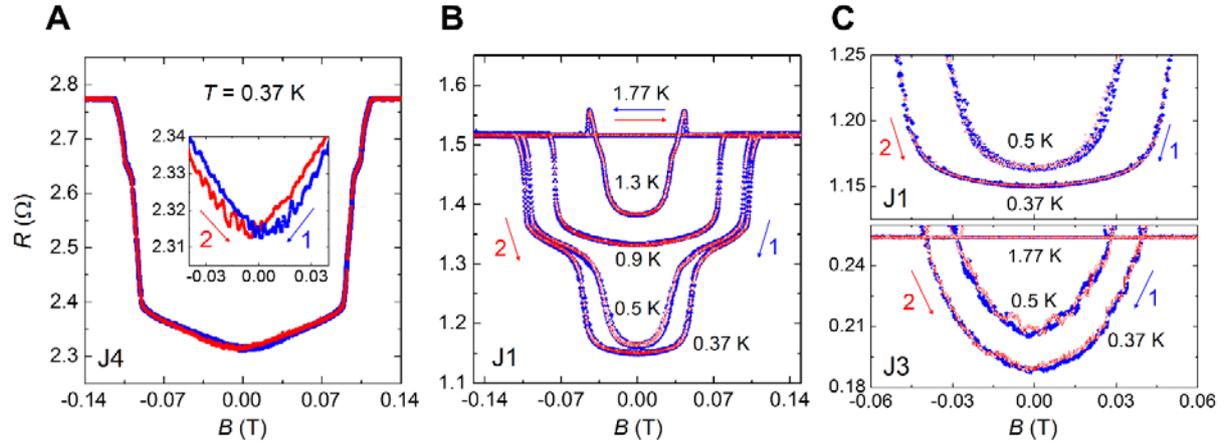

**Figure S11.** Additional zero-bias magnetoresistance data of $CoSi_2/TiSi_2$ tunnel junctions. (**A**) The magnetoresistance curves of device J4 demonstrates that the hysteresis persists down to 0.37 K (cf. the Fig. 4a in the main text). (**B**) Magnetoresistance curves of device J1 at different temperatures. (**C**) Magnetoresistance curves of devices J1 (top panel) and J3 (bottom panel) at 0.50 and 0.37 K at an enlarged scale. Note that these two devices have comparatively large S/N interface areas among the tunnel junctions studied in this work (Table S1). No hysteresis in the magnetoresistance is observed to within experimental uncertainties in these two devices. The arrows and numbers indicate the *B* sweeping sequence.

Figure S11 shows that the observed magnetoresistance hysteresis decreases as the interface area increases, as expected for random arrangements of an increasing number of chiral domains. We note in passing that a similar "advanced" magnetoresistance hysteresis was observed in Graphene Josephson junctions[48]. There, the occurrence of the hysteresis requires S/Graphene/S-type junctions and would be absent in S/N (N = Graphene) junctions of the type we are considering. Thus, Josephson junction phenomena as in Ref. 48 appear to be a rather unlikely origin of the observed magnetoresistance hysteresis.

## S12. Paramagnetic CoSi$_2$ and a CoSi$_2$/Si epitaxial film

A general concern with regard to the possible presence of magnetic clusters or phases might be a source of the measured conductance spectra and hysteretic magnetoresistances. We have carried out SQUID magnetization measurements to rule out any detectable formation of magnetically ordered phases in our CoSi$_2$/Si(100) epitaxial films.

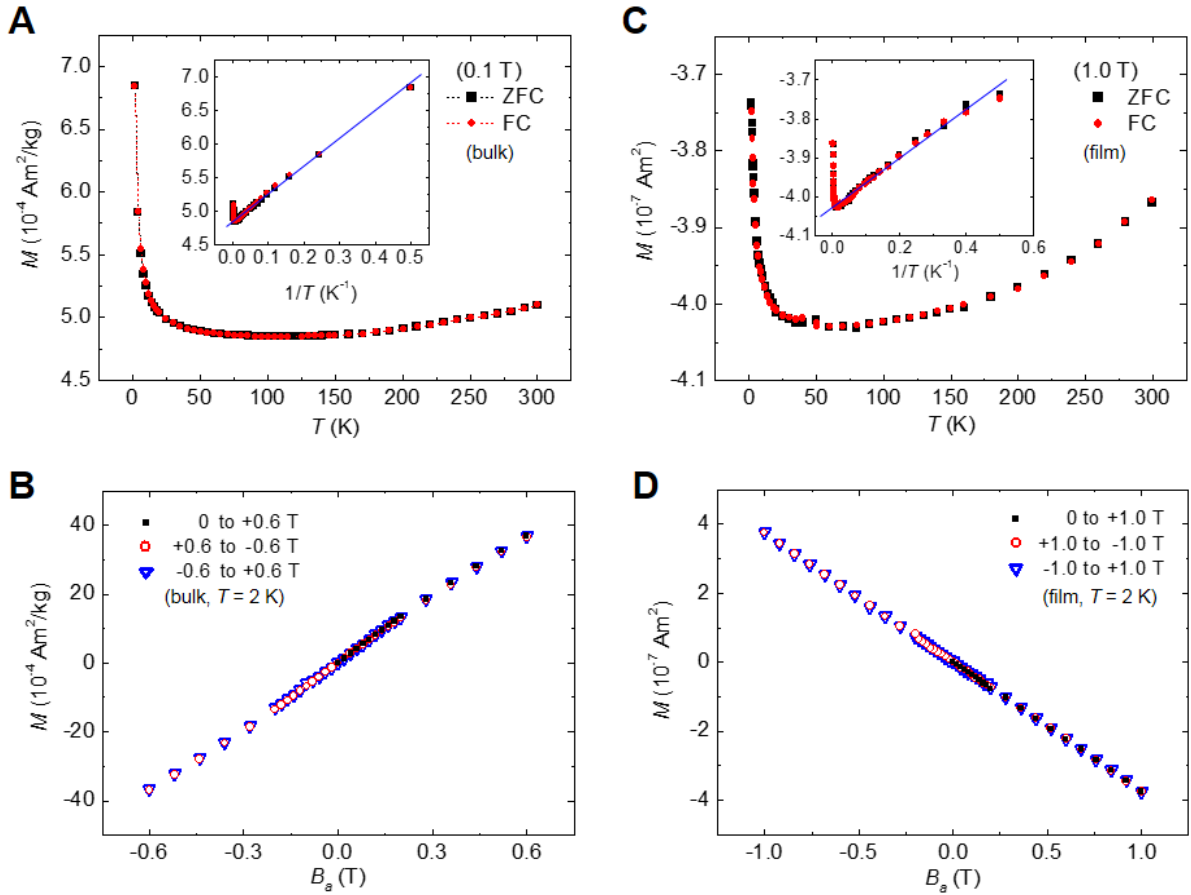

**Figure S12**. (**A**) and (**B**) The main panels show respectively the variation of mass-normalized magnetization with temperature and applied magnetic field, *i.e.*, $M(T)$ and $M(B_a)$, of a 0.196-g polycrystalline CoSi$_2$ bulk sample (99.9%, Uni-Onward Corp., Hsinchu, Taiwan). (**C**) and (**D**) The main panels show respectively the $M(T)$ and $M(B_a)$ curves for a 105-nm-thick CoSi$_2$/Si(100) epitaxial film. The negative background arises from the diamagnetic contribution of the Si substrate[49]. In Figs. (**A**) and (**C**), the magnetizations were measured through zero-field cooling (ZFC) and field cooling (FC) from 300 K down to 2 K. In both cases, the ZFC and FC results overlap. At temperatures below about 70 K, a $1/T$ dependence of the magnetization is observed (insets, where the straight lines are a guide to the eye), indicating that the samples are paramagnetic and well characterized by the Curie's law. In Figs. (**B**) and (**D**), the $M(B_a)$ curves show no hysteresis, suggesting the absence of any detectable (anti)ferromagnetic response.

## S13. Ruling out a possible contribution from the top interface via an etching experiment

A general concern with regard to the conductance spectra might be a possible contribution of the top interface to the S/N conductance spectra. We have performed a control experiment to address this point. In this experiment we removed part of the protruding TiSi$_{2-x}$ section on top of CoSi$_2$ with chemical etching, and then compared the transport properties and conductance spectra of this junction device before and after etching. As a result, we can safely rule out that the characteristic features of our S/N heterojunctions are substantially modified or even caused by the top layer.

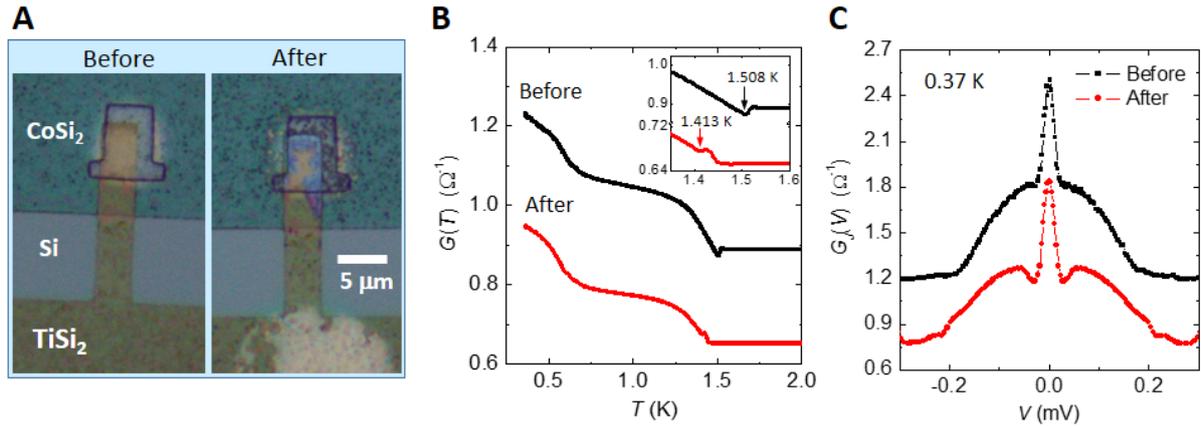

**Figure S13.** (**A**) Optical images (top view) of a CoSi$_2$/TiSi$_2$ junction device before and after etching with dilute HF aqueous solution. After etching, the dark-yellow TiSi$_{2-x}$ section inside the EBL defined window was largely removed. (**B**) The temperature dependence of zero-bias conductance $G(T)$ clearly shows that the effect of etching only results in a decrease of the $G(T)$ curve, while the overall temperature behavior remains essentially unchanged. Inset: a zoom-in around 1.5 K. (**C**) The finite-bias $G_J(V, T = 0.37 \text{ K})$ curves possess a similar lineshape before and after etching, strongly demonstrating a dominant role of the (vertical) "active" CoSi$_2$/TiSi$_2$ junction, instead of the top CoSi$_2$/TiSi$_{2-x}$ interface, in producing the conductance spectra.

**Table S1. Device parameters of CoSi$_2$/TiSi$_2$ (S/N) tunnel junctions.** $R_J$ is the junction resistance of the S/N interface. $(dI/dV)_n$ lists the magnitude of the normalized two-step zero-bias conductance. The magnitude of the broad hump is also listed (cf. Fig. 2 in main text). $\Delta$ is the superconducting gap amplitude. $Z$ is the dimensionless tunneling barrier height extracted from least-squares fits of the measured conductance spectra as described in S8. $\Delta B$ is the size of hysteresis in junction magnetoresistance defined in Fig. 4 in main text. All values are measured at 0.37 K.

| Device | $R_J$ (Ω) | $(dI/dV)_n$ | $(dI/dV)_n$ (broad hump) | $\Delta$ (meV) | $Z$ | Active junction area (μm$^2$) | $\Delta B$ (G) |
|---|---|---|---|---|---|---|---|
| J1 | 0.61 | 2.16 | 1.66 | 0.225 | 0.67 | 0.58 | 0 |
| J2 | 0.92 | 1.55 | 1.39 | 0.24 | 0.81 | 0.22 | ~170 |
| J3 | 0.16 | 2.36 | 1.36 | 0.12 | 0.1 | 1.6 | 0 |
| J4 | 0.45 | 1.99 | 1.59 | 0.14 | 0.84 | 0.10 | ~110 |
| J5 | 1.15 | 1.34 | 1.31 | 0.22 | 0.83 | 0.05 | ~400 |

**Table S2. Device parameters of T-shaped CoSi$_2$/TiSi$_2$ superconducting proximity structures.** $R_N$ ($\rho_N$) is the residual resistance (resistivity) of TiSi$_2$ wire. $L_1$ and $L_2$ are defined in Fig. 3a in the main text. $\Delta(dI/dV)_n \equiv (dI/dV)_n - 1$ is the increase in normalized conductance at 0.37 K. FWHM is the full width at half maximum of the ZBCP. The Thouless energy is given by $E_{Th} \approx \hbar D_{DN}/L_1^2$, where $D_{DN}$ is the diffusion constant of TiSi$_2$. $T_{onset}$ is the onset temperature of the proximity effect. A4$_{ren}$ lists the renormalized parameters for the A4 device, as described in S10.

| Device | $R_N$(2K) (Ω) | $\rho_N$(2 K) (μΩ cm) | $L_2$ (μm) | $L_1$ (μm) | $\Delta(dI/dV)_n$ | FWHM (meV) | $E_{Th}$ (meV) | $D_{DN}$ (cm$^2$/s) | $T_{onset}$ (K) |
|---|---|---|---|---|---|---|---|---|---|
| A1 | 0.59 | 4.9 | 0.5 | 0.24 | 0.25% | 0.032 | 0.053 | 46 | 1.44 |
| A2 | 0.77 | 3.5 | 0.07 | 0.34 | 0.16% | 0.054 | 0.037 | 65 | 1.49 |
| A3 | 1.61 | 5.4 | 0.37 | 0.42 | 0.11% | 0.042 | 0.016 | 42 | 1.48 |
| A4 | 29.61 | 126 | 0.18 | 0.45 | 0.75% | 0.35 | 0.0016 | 4.9 | 1.46 |
| A4$_{ren}$ | 2.61 | 22.1 | 0.18 | 0.27 | 9.2% | 0.024 | 0.014 | 15 | 1.46 |